\documentclass{article}

\PassOptionsToPackage{numbers, compress}{natbib}

\usepackage[preprint]{neurips_2020}

\usepackage[utf8]{inputenc} 
\usepackage[T1]{fontenc}    
\usepackage{hyperref}       
\usepackage{url}            
\usepackage{booktabs}       
\usepackage{amsfonts}       
\usepackage{nicefrac}       
\usepackage{microtype}      

\usepackage{multirow}
\usepackage{graphicx}
\usepackage{subcaption}
\usepackage{xcolor}

\title{Robust watermarking with double detector--discriminator approach}

\author{
  Marcin Plata$^{1,2}$ \\
  \texttt{marcin.plata@pwr.edu.pl}  \\
  \\
  $^{1}$ Department of Fundamentals of Computer Science \\
  Wroclaw University of Science and Technology \\
  Wroclaw, Poland \\
   \And
  Piotr Syga$^{1,2}$ \\
  \texttt{piotr.syga@pwr.edu.pl}  \\
  \\ 
  $^{2}$ Vestigit \\
  Wroclaw, Poland \\
  \\
}

\begin{document}
\maketitle

\begin{abstract}
  

In this paper we present a novel deep framework for a watermarking --- a technique of embedding a transparent message into an image in a way that allows retrieving the message from a (perturbed) copy, so that copyright infringement can be tracked. For this technique, it is essential to extract the information from the image even after imposing some digital processing operations on it. Our framework outperforms recent methods in the context of robustness against not only spectrum of attacks (e.g. rotation, resizing, Gaussian smoothing) but also against compression, especially JPEG. The bit accuracy of our method is at least $0.86$ for all types of distortions. We also achieved $0.90$ bit accuracy for JPEG while recent methods provided at most $0.83$. Our method retains high transparency and capacity as well. Moreover, we present our double detector--discriminator approach --- a scheme to detect and discriminate if the image contains the embedded message or not, which is crucial for real--life watermarking systems and up to now was not investigated using neural networks. With this, we design a testing formula to validate our extended approach and compared it with a common procedure. We also present an alternative method of balancing between image quality and robustness on attacks which is easily applicable to the framework.

\end{abstract}

\section{Introduction}

The research on watermarking techniques in the multimedia sector is significantly rising, as both the presented methods achieve ever-increasing accuracy at the same time as mitigating the flaws of embedding additional information into the image, and the need for pervasive marking of the intellectual property due to increasing rate of piracy emerges. As the multimedia services are working on simplifying the content delivery systems and broadening its accessibility, the increased number of online assets allows to easily intercept the content and redistribute it illegally to violator's own benefit, somewhat utilizing the amount of media online to hide his infringement. In order to prevent it, many techniques are used, with watermarking being one of the most effective methods~\cite{IBC,report2018}. 
The idea of digital watermarking assumes that one embeds an additional message into an image or a video, so that when copied, the owner of the content may prove his rights. Moreover, when digital media is made available to a wide range of users, the embedding of a  watermark  that is unique for each user, allows, in the case of an unauthorised redistribution, to indicate which user has leaked the content. 
The aim is to allow any authorized party easy localization, extraction and identification of the watermark, even if the copy has been manipulated.
Since a person that aims to violate the copyright wants to destroy the embedded watermark, it has to be \textit{robust} against image processing attacks that remove identification information encoded in it. Moreover, as the main interest of copyright owner is to provide the highest possible quality of his property to the clients, the embedded watermark should not be visible for the end user (\textit{transparency}), nor should it diminish the quality of the image. It is easy to note, that the requirements of transparency and robustness are to some extent opposing (assume a constant capacity), hence it is of paramount importance for a watermarking system to provide an appropriate balance between the two.
Naturally, the higher demand on the content, the better it should be protected, hence the watermark should have the \textit{capacity} that allows to identify a large number of unique end users, Note that not only adversarial attacks hinder the effectiveness of watermarking. In order to maximize the network transfer, images and videos are compressed so that almost all surplus information, i.e., the information that is not properly interpreted by human perception, is removed, leaving only the data that are important for the visual quality. An example of the information that is removed in compression (e.g. JPEG) is chrominance (Cr, Cb) for which human sight is more oblivious than the luminance (Y) of an object. Since information in chroma components of the colorspace is trimmed significantly, the watermark has to be embedded into the components that end user is more aware of. 
Note that, in simplification, video watermarking may be viewed as an extension of image watermarking with a obstacle -- MPEG compression, encoding primary frames (I) using JPEG and other frames (P and B) by relying on references to the primary. 

\paragraph{Related Work.}
Various methods has been used in order to provide robust and transparent watermarking, in~\cite{Najafi,Kumar2018} the authors used Discrete Cosine Transform (DCT) in order to comply with JPEG compression, whereas a different frequency approach, combined with Singular Value Decomposition (SVD) was proposed in~\cite{NAJAFI2019,Liu2019} (sharp frequency localized contourlet transform and discrete wave transform were used respectively) as frequency-domain modifications allow to easily 'spread' the information across the visible image. In~\cite{dual}, the authors used redundancy in their dual watermarks, to improve the robustness against cropping.
Similar effect was achieved by an end-to-end encoder-noiser-decoder framework proposed in~\cite{Zhu_2018_ECCV} that spread the embedded message over all pixels.
A follow up paper~\cite{wen2019romark} introduced adversarial training, that resulted in further improvement of the robustness, albeit at the expense of the transparency.  In fact, the resulting image significantly deviates from the well established ranges of 'acceptable' PSNR, mitigating its commercial value.
A spread spectrum watermarking with adaptive strength  and differential quantization allowed the authors of~\cite{8673925} to improve PSNR guarantees. 
The robustness of the algorithm was the focus of~\cite{luo2020distortion}, where authors add another neural network for generating generic distortions in the training. The presented model allowed to achieve improvement on previous accuracy, however the authors did not provide tests of robustness against such common attacks as rotation and subsampling. The authors of~\cite{spatial} focused on the local capacity of an image, presenting a method improving robustness of the watermark against wide range of the attacks, as well as proposing a differentiable approximation of JPEG compression. The latter was investigated also in~\cite{ReDMark}. Coping with the difficulties introduced by image compression was focused on likewise in~\cite{JpegRotate}. 
\paragraph{Our Contribution.} In this paper, we introduced (1) a novel end-to-end watermarking system utilizing an additional component to inspect if images contain a watermark. (2) We enhanced the architecture of the system and extended the encoder by a watermark adapter, resulting in a significant improvement of the robustness against some attacks and compression algorithms, such as rotation, resizing or JPEG. (3) We proposed a novel evaluation method coping with false suspects of a copyright violation and indicated the efficiency of our discriminator--detector approach. (4) We also provided an alternative method to increase the transparency of encoded images and shown its efficiency.

\section{Method}

\begin{figure}
    \centering
    \includegraphics[width=.9\textwidth]{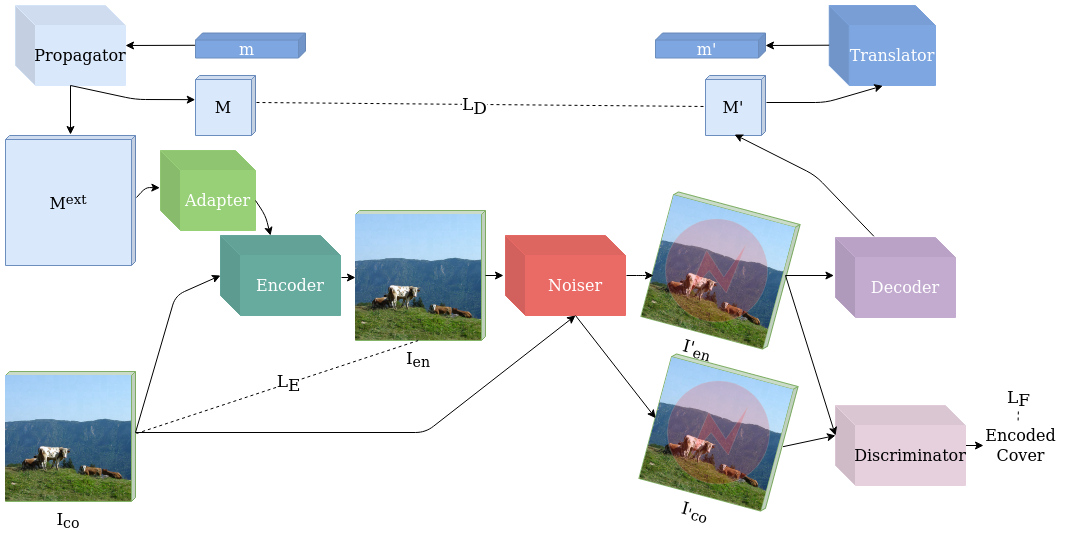}
    \caption{The schema of the pipeline architecture. The dashed lines link objects connected with the respective loss function. The red lightnings denote distortions given by the nosier.}
    \label{fig:arch}
\end{figure}

The aim of watermarking techniques is to embed some binary data \(m \in \{0,1\}^L\) into a cover image \(I_{co}\) of shape \(W \times H \times C\). To achieve that, we use the \textit{encoder} \(E\) which returns the encoded image \(I_{en}\). The encoded image needs to pass transparency requirement, i.e., be perceptually similar to the cover $I_{co}$. Moreover, the encoded image should be robust against some processing operations called \textit{attacks} (we utilize \textit{noisers} in the training phase to simulate attacks). Then, we process the distorted image \(I'_{en}\) using the \textit{decoder} \(D\) to extract the message \(m' \in \{0,1\}^L\). We aim to receive \(|m' - m| < \delta\), i.e. the extracted message should be similar to the embedded one. Note that, in the real--life scenario we also need to determine if the investigated image contains the additional data or not. To distinguish between distorted images \(I'_{co}\) and \(I'_{en}\), we use the \textit{discriminator} \(F\). \\
Our system contains eight components, where three of them are trainable neutral networks ---  \textit{encoder} \(E\), \textit{decoder} \(D\) and \textit{discriminator} \(F\). Next two are differentiable layers - \textit{noiser} \(N\) and \textit{prenoiser} \(N_{\mathrm{pre}}\), used for adding artificial noise to the images. The last two elements, the \textit{propagator} \(P\) and \textit{translator} \(T\),  are required to propagate the message to a spatial form and revert this operation. We also distinguished an auxiliary neural network called \textit{adapter} \(A\) which extends the propagator \(P\). The overall architecture was presented in Fig.~\ref{fig:arch}.

\paragraph{Propagator, adapter and translator.}
In our solution, we applied a method to reduce local \textit{bits-per-pixels} capacity.  Instead of assigning a representation of the message \(m\) to every pixel of the cover image \(I_{co}\), as in \cite{Zhu_2018_ECCV,luo2020distortion}, among others, we used a spatial spreading algorithm proposed in~\cite{spatial}. The algorithm converts the message \(m\) to a sequence \(s = (s_0, s_1, \dots, s_{\lceil \frac{L}{k} \rceil-1})\), where \(s_i\) is a tuple containing a slice of the message \(m_{[ik, \dots, ik+k-1]}\) and a reference index in a binary form \(\mathrm{bin}(i)\). We assumed that \(k\) is a size of the slice and \(i \in \{0, \dots, \lceil \frac{L}{k} \rceil -1 \} \). It is easy to notice that, we need at most \(\lceil \log_2(\frac{L}{k}) \rceil \) bits to store the binary index \(\mathrm{bin}(i)\). For example, for \(k=2\) and \(L=32\), we need to prepare a sequence \(s\) with length equal to \(16\) and a tuple \(s_i\) containing 6 bits (4 bits for \(bin(i)\) and 2 bits for \(m_{[i,i+1]}\)). We also define \(k' = k + \lceil \log_2(\frac{L}{k}) \rceil \) for simplicity of further formulations. \\   
The propagator \(P\) converts the message \(m\) to a sequence \(s\) and generates two spatially-spread representations of the message, i.e. \(M \in \{0,1\}^{\frac{W}{b} \times \frac{H}{b} \times k'}\) and \(M^{ext} \in \{0, 1\}^{W \times H \times k'}\), where \(b\) is an argument of the propagator, which refers to a size of a unitary block of a size \(b \times b \times k'\), which contains some tuple \(s_i\) replicated \(b\) times in two directions. The tuples \(s_i\) and the unitary blocks are randomly sampled in order to fill the messages \(M\) and \(M^{ext}\), respectively. In the case of \(M\), we assign 
tuples \(s_i\) to cells of a grid of a shape \(\frac{W}{b} \times \frac{H}{b}\), while in the case of \(M^{ext}\), to every cell of the grid we assign a unitary block of shape \(b \times b \times k'\), thus the final shape of \(M^{ext}\) is equal to \({W \times H \times k'}\). The translator \(T\) works with the message \(M\) and reverts the operations applied by the propagator \(P\). In order to extract some slice of the message \(m\), the translator chooses \(n\) tuples (cells) in which stored binary indices are closest to a corresponding index of the slice, then elements of the chosen tuples referring to the slice of the message are averaged. \\
We extended the propagator \(P\) by adding the adapter \(A\), which is a convolutional neutral network used to adapting the propagator output to a convenient form for the encoder \(E\). Similar approach was presented in \cite{luo2020distortion}, where linear layers were used to adapt the message \(m\) directly. The adapter \(A\) was separated from the encoder \(E\) as it allowed to produce the result of \(A(M^{ext})\) independently of the processing steps of the images. It could be particularly relevant in a case of working with a sequence of frames. The adapter \(A\) is build with four convolutional blocks called \textit{conv-bn-relu}, which contains the convolutional layer, batch normalization and ReLU activation. The last block outputs 6 channels.

\paragraph{Discriminator.} 

The primary role of the discriminator \(F\) was the application of \textit{adversarial training} approach \cite{gan,steggan} in order to improve the perceptual similarity of the encoded and cover images. We also utilized it to indicate if the image contains the message or not. 
The details and motivation to use the discriminator in this way were presented in Sect.~\ref{sec:double}. The discriminator is build with 3 conv-bn-relu blocks, a global average pooling layer, a linear layer with single output unit and Sigmoid activation.

\paragraph{Encoder and decoder.}

The encoder \(E\) ands decoder \(D\) are two main components of our watermarking system.  Both networks need to cooperate during training in order to find the balance between the transparency and robustness requirements of the watermarking. Moreover, they determine a joint "scheme" of the encoding--decoding procedure which is known only for the trainable components of the pipeline. The watermarking system is designed to work with some real--life limitations. 
As an example, note that watermarking videos is currently in much higher demand than watermarking strictly still images, hence any proper image watermarking method should be easily extended for marking movies (e.g., at frame by frame basis). In the \textit{over-the-top} (OTT) media, services providing video--streaming are severely constrained by the content servers (origin or cache) that, due to storage limitations, are not able to provide different contents to all users (with a key per user).  As a consequence, to allow uniquely watermarked media for each user, the encoder needs to work on the client's side and handle a high quality video~\cite{mts}. This indicates that the proposed architecture of the encoder (and indirectly other neural networks) has to be relatively small and shallow. Additionally, it explains the reason of considering the adapter \(A\) as the separated component, which allowing to be used once for all frames.  The encoder processes the cover image using three conv-bn-relu blocks. Next, the output is concatenated with the cover images and the adapted message taken from \(A\), followed by the application of two conv-bn-relu blocks. Finally, we use the convolutional layer with the kernel size equal to \(1\) on the concatenation of the cover image and the prior output. 
The detector is based on eight conv-bn-relu blocks. We apply average pooling with the kernel size and stride equal to \(b\) to the output. Finally, we use the conv-bn-relu and the convolution with the kernel sizes equal to \(1\) for both. 

\paragraph{Attacks and noiser layers.} 

In our work, we considered eight attack types, all implemented in the noisier layer \(N\). The \textit{crop} attack crops a square form the encoded image and it is parameterized by \(p\) referring to a ratio of the squared area. The \textit{cropout} also crops the square while the rest of the images is replaced by the cover image. The \textit{dropout} attack chooses pixels with the probability \(p\) and replaces them with the corresponding pixels of the cover image. Both cropout and dropout imitate binary symmetric channel which in information theory describes more challenging model of communication than binary erasure channel (that may be simulated by \textit{salt and pepper} noising) \cite{Zhu_2018_ECCV}. We also included common computer vision operations -- resizing, rotation and Gaussian smoothing, as well as most common compression algorithm  -- JPEG (with quality parameter \(q\)) and 4:2:0 chroma subsampling procedure. The JPEG algorithm includes \textit{non-differentiable} operations, e.g. quantization, thus we could not apply it in the training pipeline not halting the neural network weights updates. To handle this problem, we used an approximation of the JPEG proposed in \cite{ReDMark, spatial}. In the experiments described in Sect.~\ref{sec:qual}, the noiser layer \(N_{\mathrm{pre}}\) is executed before \(N\) and always applies the dropout. 

\paragraph{Training details and hyper parameters.}

To train and test our method, we used the COCO dataset~\cite{coco}. We sampled $10000$ images for the training subset and $1000$ for the testing subset. Both subsets were disjoint. We resized images to \(256 \times 256\) pixels and encoded the message of the length \(L=32\). The messages were sampled at random, the spatial spreading was random as well. We used Adam~\cite{adam} with learning rate equal to $0.001$(other parameters were set to default). The pipeline was trained with the batch size $12$ for $100$ epochs and all attacks were applied (one type in each iteration of the training). At the end, we froze all the weights except the discriminator and ran the discriminator's part of the pipeline for 20 epochs. To train the system, we used two GPUs -- Nvidia RTX 2080Ti 11GB. The one epoch of the pipeline training took about 370 seconds, while during the inference we were able to process about 45 images per second by one component and using one GPU.
The default parameters for the convolutional layers were as follow: the channel size $64$, kernel size $3$ and the reflection padding applied. For the linear layers we set $64$ units by default. All neural network were fed by images in the YCrCb color space, the same space were used for the images returned by the encoder.

\paragraph{The training procedure and objectives.}
In one iteration of the pipeline, first, we take a message \(m\) and apply it to the propagator \(P\), which returns two variants of the spatially-spread messages, i.e.
\( M^{ext}, M = P(m).\)
Next, we use the adapter \(A\) to transform the message \(M^{ext}\) and encode the cover image 
$I_{en} = E(I_{co}, A(M^{ext}))$. 
The output of the operation is the encoded image \(I_{en}\), which is distorted by the nosier layers \(N\) afterward. Exactly the same distortion needs to be applied on the cover image \(I_{co}\), thus we have:
\(I'_{en}, I'_{co}, M = N(I_{en}, I_{co}, M)\).
Note that, some types of attacks require the cover image \(I_{co}\) in order to distort the encoded image \(I_{en}\). Moreover, for the types of noises which affect the encoded image spatially, we also calibrate the message \(M\), e.g. for cropping attacks, we also crop the message \(M\). 

The decoder \(D\) is fed by a noised encoded image \(I'_{en}\) and predicts the encoded message 
${M' = D(I'_{en})}$,
while the discriminator \(F\) distinguishes between the noised cover and encoded images, \(I'_{co}\) and \(I'_{en}\) respectively, i.e., 
$F(I' \in \{I'_{co}, I'_{en}\}) \in [0,1]~.$\\
We need to determine a loss component in order to train the discriminator used for two purposes --- improving the perceptual similarity and, as an auxiliary component of our watermarking system for double discriminator--decoder approach. Naturally, the main focus is to keep the transparency and robustness at highest possible level.\\
We seek to ensure the transparency using the \textit{mean square error} between \(I_{co}\) and \(I_{en}\), thus
$L_E(I_{en},I_{co}) = \frac{1}{WHC}||I_{en}-I_{co}||^2_2$,
where \(||\cdot||_2\) is the Frobenius norm. In order to handle the message decoding, we used the \textit{mean-variance} approach proposed in~\cite{spatial}, given by:
\[L_D(M,M') = \frac{b^2}{WH} \sum_{w=0}^{W/_{b}} \sum_{h=0}^{H/_{b}} \lambda_D^{\mathrm{mean}}\mathrm{mean}(| M_{hw} - M'_{hw} |) + \lambda_D^{\mathrm{var}}\mathrm{var}(| M_{hw} - M'_{hw} |)~,\]
where \(W/_{b} = \frac{W}{b}-1\), \(H/_{b} = \frac{H}{b}-1\) and \(|\cdot|\) returns element-wise absolute values. This way of the loss formulation converges to a case in which some tuples \(M'_{ij}\) contain possibly high quality of the overal data, instead of returning the correct predictions for a proper subset of the indices for all tuples. Note that, due to high redundancy of tuples in \(M'\), this way of convergence is advisable.\\
We provided the adversarial training of the encoder \(E\) using the discriminator \(F\). The aim of the encoder is to produce an image recognized as the cover image by the discriminator, while in our pipeline the image is further distorted by the noiser \(N\). Thus, we defined \(L_E^F(I'_{en}) = \log(F(I'_{en}))\). The aim of the discriminator is to distinguish between the distorted cover and encoded images, which was formulated as \(L^F_F(I'_{co}, I'_{en}) = \log(F(I'_{co})) + \log(1-F(I'_{en}))\).\\
To obtain the parameters for the encoder \(E\), adapter \(A\) and decoder \(D\), we minimize the objective over the distribution of images and messages, namely $\mathbb{E}_{I_{co},M}[\lambda_{E}L_{E} + \lambda_{F}L_F^E + L_D]$,
with \(\lambda\) weights.
To train the discriminator \(F\), we minimize the objective over the distribution of images: \(\mathbb{E}_{I_{co}}[L_F^F]\).

\section{Watermark robustness}
\label{sec:wmrobust}
\begin{figure}
    \centering
    \includegraphics[width=\textwidth]{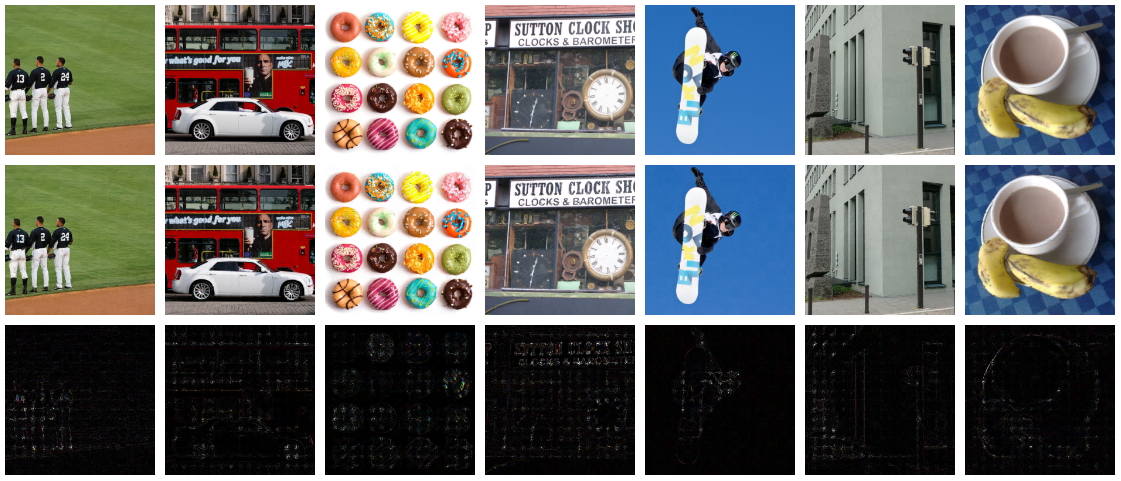}
    \caption{The cover images \(I_c\) (above), the encoded images \(I_e\) (middle) and their differences with the min-max normalization (below). 
    }
    \label{fig:imgs}
\end{figure}

In this section we present the efficiency of the image encoding--decoding procedure of our watermarking solution. In order to compare our framework with the well established ones, we applied the common evaluation approach, which relies on calculating the bit accuracy of the detected messages.\\
We set the propagator's parameters to \(k=2\) and \(b=16\) and the translator parameter \(n=3\). We split the message \(m\) into \(16\) tuples and further replicate them to fill the unitary blocks of the size \(16 \times 16 \times 6\). The expected number of the blocks' redundancy was equal to \(16\). We trained the pipeline with the following parameters: \(\lambda_E = 2.4\), \(\lambda_D^{\mathrm{mean}}=1.0\), \(\lambda_D^{\mathrm{var}}=1.0\) and \(\lambda_F=0.05\).
The results' comparison of the bit accuracy between our solution and the state-of-the-art works was presented in~Tab.~\ref{table:res}. The results were achieved for the PSNR equal to \(31.57\)dB, \(44.04\)dB and \(44.90\)dB for the Y, Cb and Cr channels, respectively. The examples of the encoded images were presented in Figure~\ref{fig:imgs}. Our solution exhibited significant improvement for the rotation, resizing and JPEG attacks. Moreover, it provided the highest, among all methods, overall accuracy, which was measured as the lowest bit accuracy of all attacks. For our solution, the lowest bit accuracy was equal to \(0.86\), while it was \(0.831\) for~\cite{spatial} and below \(0.7\) for others methods.

\begin{table}
    \begin{center}
    \caption{The robustness against selected attacks  for our method and state-of-the-art systems with respect to average of the bit accuracy. Note that the resizing modes were not declared in \cite{luo2020distortion,ReDMark}.}
    \label{table:res}
    \centering
    \begin{tabular}{c c c c c c }
    \toprule
    \multirow{2}{*}{Attacks} & \multicolumn{5}{c}{Methods} \\
     \cmidrule(r){2-6}
    & Our & Spatial \cite{spatial} & HiDDeN \cite{Zhu_2018_ECCV} & DADW \cite{luo2020distortion} & RedMark \cite{ReDMark} \\
    \midrule
    No attack & 1.000 & 1.000 & 1.000 & 1.000 & 1.000 \\
    Crop(\(p=0.3\)) & 0.860 & 0.832 &  \textbf{1.000} & \textbf{1.000} & - \\
    Cropout(\(p=0.3\)) & 0.921 & 0.902 & \textbf{0.940} & - & 0.925 \\
    Dropout(\(p=0.5\)) & 0.981 & 0.962 & \textbf{1.000} & \textbf{1.000} &  \(\approx\)0.99 \\
    Rotate(\(\alpha=5^{\circ}\)) & \textbf{0.956} & 0.842 & - & - & - \\
    Gaussian(\(\sigma=2\)) & \textbf{0.991} & 0.986 & 0.960 & 0.600 & 0.5 \\
    Gaussian(\(\sigma=4\)) & \textbf{0.989} & 0.982 & 0.820 & 0.500 & 0.5 \\
    Subsampling(4:2:0) & \textbf{0.993} & 0.984 & - & - & - \\
    Resize(\(s=0.5\), \(m=N\)) & \textbf{0.913} & 0.849 & - & \multirow{2}{*}{0.671} &  \multirow{2}{*}{0.819} \\
    Resize(\(s=0.5\), \(m=L\)) & 0.886 & \textbf{0.908} & - &  & \\
    JPEG(\(q=50\)) & \textbf{0.902} & 0.831 & 0.670 & 0.817 & 0.746 \\
    \bottomrule
    \end{tabular}
    \end{center}
\end{table}
\section{Double discriminator--decoder approach}
\label{sec:double}
In the real--life environment, watermarking systems are used only in small subset of the total multimedia content worldwide. Thus, in order to detect the message, we need to handle one of following approaches:
\begin{enumerate}
    \item na\"ive ---  apply the detection procedure on every image and label the suspects, when the extracted key shares at least \(t\) bits with any of the keys from the database,
    \item double --- at first execute a procedure to distinguish if given content comes from our watermarked sources or not
\end{enumerate}

The former relies on using a highly effective detection procedure. For example, let us assume that \(t = 29\) (estimated for accuracy about 90\%) and the watermark contains 32 bits. Having one million random keys in the database, the probability of the event that at least one key from the database contains at least 29 shared bits is equal to \(1~-~(\sum_{i=0}^{28} [{32 \choose i} 0.5^i 0.5^{32-i}])^{10^6} \approx 0.72 \). Thus, the chance of failure is high even for relatively small database of the keys and high accuracy of the decoder.\\
The latter could significantly improve the overall efficacy of the watermarking system, but it requires auxiliary subsystem to distinguish between the content sources. Instead of using the critic \(F\) only to rate if the encoded image is similar to the original one, we moved its position in the system and placed it after the nosier \(N\) in the training pipeline. By this, we fed the critic \(F\) with noised images \(I_{no}\). Therefore, cover images \(I_{co}\) also needed to be processed by the nosier \(N\) in the same way as encoded images \(I_{en}\) in order to avoid to learn inappropriate characteristics of the cover and encoded images, i.e. the critic would be able to learn features which are effects of processing by the nosier rather than the encoder. This modification still gave us possibility to use the critic \(F\) for \textit{adversarial training} and aiming to improve the transparency of the encoded images.

\paragraph{Used metrics.} \label{testing}
During the tests, we assume that an image \(I\) contains the watermark if \(F(I) > t_F\), where \(t_F \in [0,1]\) is a threshold for the critic's outputs. We utilized the standard metrics:  \textit{true positive} (TP), when \(I\) is an encoded image and \(F(I) > t_F\); \textit{false positive} (FP), when \(I\) does not contain embedded message and \(F(I) > t_F\); \textit{true negative} (TN), when \(I\) does not contain embedded message and \(F(I) \leq t_F\); \textit{false negative} (FN), when \(I\) is an encoded image and \(F(I) \leq t_F\).
While designing the watermarking system we aim at maximizing the detection of copyright infringements as well as minimizing the probability of false accusation of piracy. Following these conditions, we proposed two rates to measure the efficacy of the discriminator--decoder approach that translate well to real--life validation of the watermarking frameworks: (1) \textit{true identification rate} ($\mathrm{TIR}$), defined as a probability of extracting appropriate key provided true encoded image, i.e., \(\Pr(\mathrm{extract~true~key}|\;\mathrm{TP})\), (2) \textit{false identification rate} ($\mathrm{FIR}$), defined as a probability of indicating a wrong key, provided true encoded image or falsely classifying image as encoded, i.e., \(\Pr(\mathrm{true~key~not~extracted}|\;\mathrm{TP})\Pr(I_{en}) + \Pr(\mathrm{FP})\Pr(I_{co})\). It is easy to note that $\mathrm{TIR}$ covers the maximization problem discussed above, whereas $\mathrm{FIR}$ complies with the minimization. Working with both rates gives us ability to precisely validate the watermarking system. We also consider \(\mathrm{FIR}\) for false identification of encoded images \(I_{en}\) and cover images \(I_{co}\) separately. Hence, we define \(\mathrm{FIR}_{en} = \Pr(\mathrm{true~key~not~extracted}|\;\mathrm{TP})\) to test an indication of wrong keys from encoded images, and \(\mathrm{FIR}_{co} = \Pr(\mathrm{FP})\) to validate falsely classified cover images.

\paragraph{Robustness of the discriminator--decoder approach.}

To the best of our knowledge, recent work on neural networks-based watermarking did not include experiments following similar conditions to those described in Sect.~\ref{testing}. Vast majority of the research focuses on improving only the detection procedures and transparencies of encoded images. We validate our system using the common in recent works procedure, e.g. \cite{spatial,Zhu_2018_ECCV,luo2020distortion,ReDMark} (see Sect.~\ref{sec:wmrobust}), as well as $\mathrm{TIR}$ and $\mathrm{FIR}$. \\
Taking advantage of this testing formula, we shown the efficiency of our discriminator--decoder approach compared to the na\"ive system with a single decoder. The experiments were done for $1000$ images. We assumed that the size of the keypool is equal to \(10^6\). The results are shown in Table~\ref{table:double}. The threshold was adjusted to obtain the highest \(\mathrm{TIR}\) for the na\"ive approach, however, due to high sensitivity of \(\mathrm{FIR}_{co}\), we chose lower one if the difference between them was at most \(5\%\). Then, we adjusted the rates for the double approach keeping the similar \(\mathrm{TIR}\).  
Our results confirmed the high efficiency of the double approach. In most cases, the obtained results were higher than in the na\"ive one, notably for some attacks, the difference was significant. The double approach improves the overall performance significantly for some of the attacks in which the decoder presents relatively low bit accuracy, i.e., lower than \(0.95\). For example, having similar values for \(\mathrm{TIR}\), we completely discarded \(\mathrm{FIR}_{co}\) from around \(0.7\) and reduced \(\mathrm{FIR}_{en}\) around \(2\)-\(5\) times for cropping and cropout attacks. In turn, we observed that the na\"ive approach is as efficient as the double approach if the decoder demonstrates a high robustness against some attacks. Moreover, we observed that having inefficient discriminator and robust detector, the results could be better for the na\"ive approach, e.g., for Gaussian smoothing, the decoder's bit accuracy was close to \(0.99\) and the discriminator's -- \(0.8\) (on balanced data) and finally the overall performance was higher for the na\"ive approach. 

\begin{table}
    \begin{center}
    \caption{The results of the robustness against some attacks validated using our novel testing formula. The notably better scores are presented in bold.}
    \label{table:double}
    \centering
    \begin{tabular}{c | c c c | c c c}
    \toprule
    \multirow{2}{*}{Attacks} & \multicolumn{3}{c |}{Double} &  \multicolumn{3}{c}{Na\"ive} \\
    \cmidrule(r){2-7}
    & \(\mathrm{TIR}\) & \(\mathrm{FIR}_{en}\) & \(\mathrm{FIR}_{co}\) & \(\mathrm{TIR}\) & \(\mathrm{FIR}_{en}\) & \(\mathrm{FIR}_{co}\)  \\
    \midrule
    No attack & 0.999 & 0.001 & 0.000 & 0.999 & 0.001 & 0.005 \\
    Crop(\(p=0.3\)) & 0.219 & \textbf{0.122} &\textbf{0.000} & 0.219 & 0.594 & 0.707 \\
    Cropout(\(p=0.3\)) & 0.550 & \textbf{0.193} & \textbf{0.001} & 0.556 & 0.372 & 0.720 \\
    Dropout(\(p=0.5\)) & 0.940 &0.007 & 0.008 & 0.951 & 0.004 & 0.006 \\
    Rotate(\(\alpha=5^{\circ}\)) & 0.983 & 0.016 & \textbf{0.000} & 0.979 & 0.009 & 0.119 \\
    Gaussian(\(\sigma=2\)) & 0.950 & 0.000 & 0.474 & 0.998 & 0.000 & \textbf{0.130} \\
    Gaussian(\(\sigma=4\)) & 0.956 & 0.001 & 0.410 & 0.999 & 0.001 & \textbf{0.005} \\
    Subsampling(4:2:0) & 0.971 & 0.000 & 0.000 & 0.975 & 0.000 & 0.001 \\
    Resize(\(s=0.5, m=N\)) & 0.630 & \textbf{0.119} & \textbf{0.602} & 0.645 & 0.284 & 0.717 \\
    Resize(\(s=0.5, m=L\)) & 0.540 & \textbf{0.126} & 0.701 & 0.542 & 0.371 & 0.707 \\
    JPEG(\(q=50\)) & 0.712 & 0.069 & \textbf{0.000} & 0.712 & 0.044 & 0.118 \\
    \bottomrule
    \end{tabular}
    \end{center}
\end{table}

\section{Transparency versus robustness}
\label{sec:qual}


One of the most challenging problems of watermarking techniques is encoding message into an image in transparent manner that enables accurate decoding of the message from a distorted image. A robust watermarking needs to store the message in a part of the image, that is the most resistant on distortions resulting from attacks as well as compression algorithms.\\ 
Finding a suitable trade--off between those ratios is possible by a proper tuning of hyper parameters of the loss function during the training of the pipeline. Despite the fact that the tuning for our pipeline is time--consuming and exhaustive process, including changing parameters in the middle of the training procedure, the pipeline of three neural networks was usually not balanced the way that was expected. Additionally, in a case of creating a system that produces images with different transparency levels, the basic approach requires training separated neural networks. Thus, we extended our framework by a component which allows improving the quality of the encoded image.\\
We designed a method to improve transparency of the encoded image that could be applied after the training process. The method performs following steps:
\begin{enumerate}
    \item select a mask \(S \in \{0,1\}^{W \times H \times C}\), where \(S_{whc} \sim \mathcal{B}(p)\), i.e., sampling from Bernoulli distribution with the parameter \(p\),
    \item update \(I_{en} \leftarrow S \circ I_{en} + (1-S) \circ I_{co} \), where \(\circ\) denotes element-wise multiplication.
\end{enumerate}


\paragraph{Results.}

\begin{figure}
    \centering
   \begin{subfigure}[b]{0.4\textwidth}
        \includegraphics[width=\linewidth]{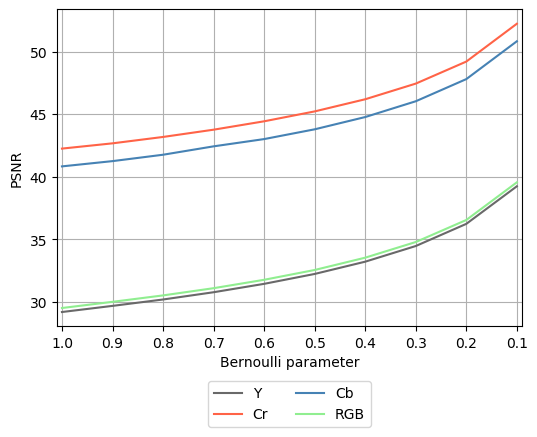}
        \subcaption{PSNR for the Bernoulli parameter \(p\).}
    \end{subfigure}
    \hfill
        \begin{subfigure}[b]{0.5\textwidth}
        \includegraphics[width=\linewidth]{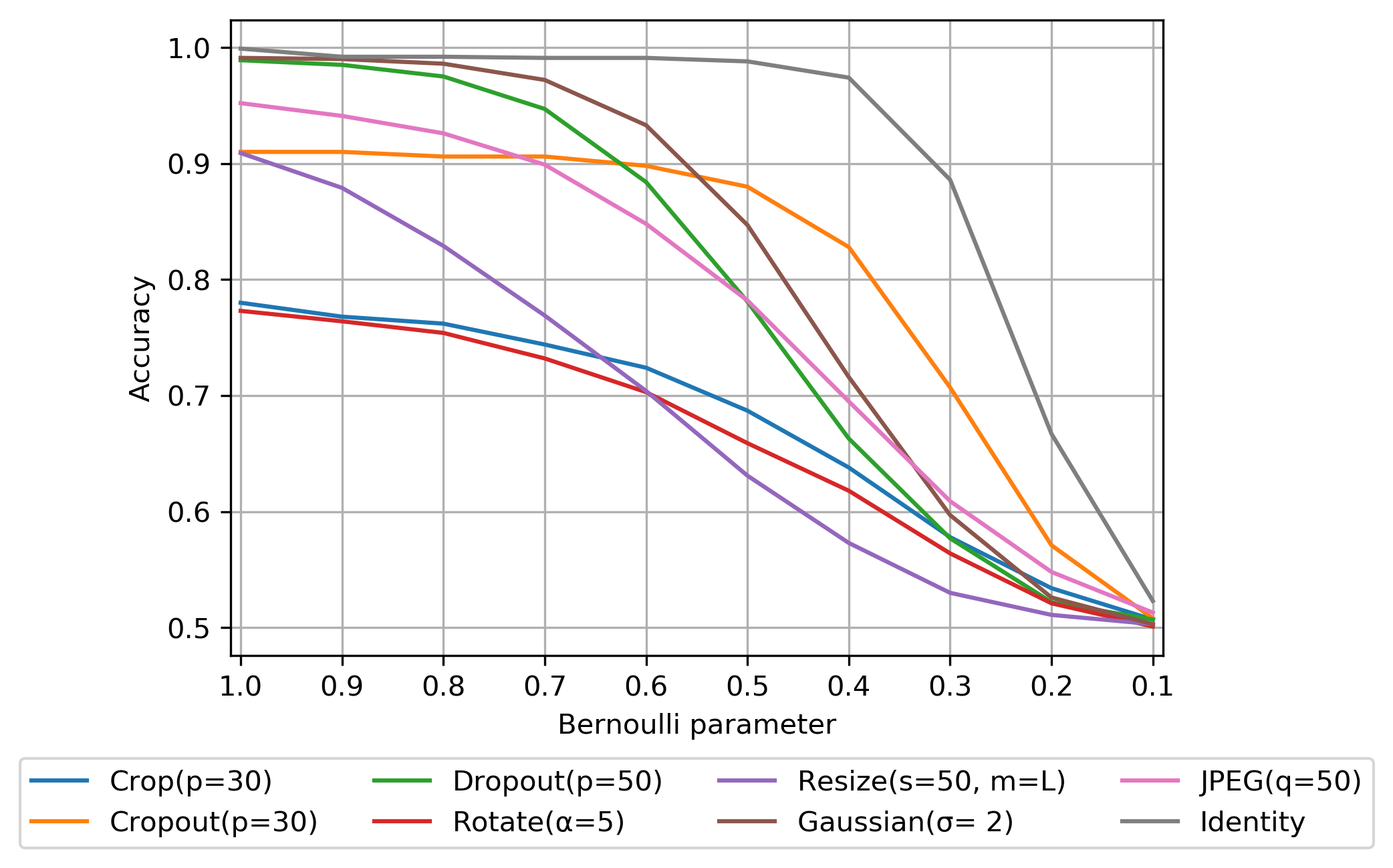}
        \subcaption{Bit accuracy for the Bernoulli parameter \(p\).}
    \end{subfigure}
    \caption{The visualization of the experiment of improving the transparency of images using dropout.}
    \label{fig:droput}
\end{figure}

We trained our pipeline with the hyper parameters set in a firmly favourable way for improving robustness against attacks, i.e. 
we set \(\lambda_E = 3.0\), \(\lambda_D^{mean}=1.0\), \(\lambda_D^{var}=1.0\) and \(\lambda_F=0.01\). Then, we compared the encoded image quality and the robustness against attacks for various values of the parameter \(p\) applied to the Bernoulli distribution. The results was presented in Figure~\ref{fig:droput}. The results confirmed that an increase of the images' transparency has negative impact on the robustness against some attacks. In \cite{ReDMark}, authors presented another method of increasing the transparency by calculating linear combination of the cover and encoded images. Note that, their system is not designed with consideration of spatial attacks (e.g. rotation), whereas we presented the method which handles this type of attacks as well.

\section{Conclusion}

In our work, we introduced a novel end-to-end watermarking solution based on neural networks. We significantly improved the robustness against some attacks, e.g. JPEG compression, rotation. The method stands out with the highest overall accuracy which is equal to \(0.86\). It is also characterized by one of the highest transparencies of the encoded images. We added a new component called \textit{adapter} to the pipeline and used a novel architecture of the pipeline in which we were able to utilize the discriminator to distinguish between cover and encoded images. We proposed an evaluation method which fits to the real--life environment of the watermarking system based on three metrics, namely \(\mathrm{TIR}\), \(\mathrm{FIR}_{en}\) and \(\mathrm{FIR}_{co}\). We evaluated a na\"ive watermarking system and our double decoder--discriminator architecture following our evaluation method. It confirmed the high efficiency of our double approach. Finally, we explored the problem of \textit{robustness versus transparency} of the watermarking systems and proposed a flexible solution to handle the problem. In the future work, we would like to continue to improve the robustness of the watermarking system, including multi-attack scenarios and new types of attacks. Moreover, we would like to enhance the capacity of the watermark as well as apply other quality measures in the training pipeline that could improve the transparency of the encoded images.

\section*{Broader Impact}

This paper presents a complete framework of image watermarking that allows both embedding and identifying the message encoded in the image. The natural environment for implementing the solution is copyright management. Due to high accuracy and low impact on the image quality, the solution may be used to protect intellectual property in the form of professional or amateur pictures, computer generated graphics or movies. Encoding a watermark may allow to identify the author of the intellectual property as well as help to identify the person that leaked the medium, allowing the breach of the copyright. For the latter case, it is required that the watermark is unique for each client.
An additional branch that may benefit from the watermarking technique is emerging as a response to the needs of deep learning solutions, data annotation business.  
Both above cases present a scenario where content creators (or owners) are protected against copyright infringement and may prove their authorship or allows presenting proof in legal actions. Naturally, such actions may be taken only in the case of high certainty. Despite the high accuracy of the method it is not foolproof, hence it may give a false positive identification and has to be sufficiently supervised so that false accusation is not made. Aside from the intended use, the popularity of watermarking may incur some negative aspects, as the legal system adjusts to the proofs by presenting the watermark we may be witnessing a 'copyright trolling' similar to recent patent rivalries between companies as some individuals may try to watermark their yet-not encoded-content with their signature, to claim ownership. This aspect may be enhanced by the false positive identification mentioned above.
The watermarking systems are a proposition for the multimedia industry to protect their businesses against the piracy. The solution considers many aspects required for solutions working in the real--life environment. The proposed solution is flexible in a context of the size of the image. It is a 'lightweight solution' as its time performance is relatively high, i.e. 30 FPS movie could be encoded in a real time harnessing graphic processing unit (GPU). The solution is also robust against video compression algorithms (most informative units of a compress movie, IFrames, are encoded using JPEG algorithm).



\small

\bibliographystyle{unsrtnat}
\bibliography{refs}

\begin{thebibliography}{19}
\providecommand{\natexlab}[1]{#1}
\providecommand{\url}[1]{\texttt{#1}}
\expandafter\ifx\csname urlstyle\endcsname\relax
  \providecommand{\doi}[1]{doi: #1}\else
  \providecommand{\doi}{doi: \begingroup \urlstyle{rm}\Url}\fi

\bibitem[{IBC's website}()]{IBC}
{IBC's website}.
\newblock {Using forensic watermarking to protect UHD content}.
\newblock
  \url{https://www.ibc.org/publish/using-forensic-watermarking-to-protect-uhd-content/946.article},
  last accessed on 2020-04-20.

\bibitem[Thorwirth et~al.(2018)Thorwirth, Hietbrink, Haitsma, Doërr, Deen,
  Wilkinson, Wilson, Stevenson, and White]{report2018}
Niels Thorwirth, Erik Hietbrink, Jaap Haitsma, Gwenaël Doërr, Glenn Deen,
  Mike Wilkinson, Robin Wilson, Brian Stevenson, and Christopher White.
\newblock {Forensic Watermarking Implementation Considerations for Streaming
  Media}.
\newblock \emph{{Streaming Video Alliance}}, July 2018.

\bibitem[{Najfi}(2017)]{Najafi}
E.~{Najfi}.
\newblock A robust embedding and blind extraction of image watermarking based
  on discrete wavelet transform.
\newblock In \emph{Mathematical Sciences}, volume~11, pages 307--318, Dec 2017.
\newblock \doi{10.1007/s40096-017-0233-1}.

\bibitem[Kumar et~al.(2018)Kumar, Singh, and Kumar]{Kumar2018}
Chandan Kumar, Anuj~Kumar Singh, and Priyadarshni Kumar.
\newblock Improved wavelet-based image watermarking through spiht.
\newblock \emph{Multimedia Tools and Applications}, pages 1--14, 2018.

\bibitem[Najafi and Loukhaoukha(2019)]{NAJAFI2019}
E.~Najafi and K.~Loukhaoukha.
\newblock Hybrid secure and robust image watermarking scheme based on svd and
  sharp frequency localized contourlet transform.
\newblock \emph{Journal of Information Security and Applications}, 44:\penalty0
  144 -- 156, 2019.
\newblock ISSN 2214-2126.
\newblock \doi{https://doi.org/10.1016/j.jisa.2018.12.002}.
\newblock URL
  \url{http://www.sciencedirect.com/science/article/pii/S2214212618302990}.

\bibitem[{Liu} et~al.(2019){Liu}, {Huang}, {Luo}, {Cao}, {Yang}, {Wei}, and
  {Zhou}]{Liu2019}
J.~{Liu}, J.~{Huang}, Y.~{Luo}, L.~{Cao}, S.~{Yang}, D.~{Wei}, and R.~{Zhou}.
\newblock An optimized image watermarking method based on hd and svd in dwt
  domain.
\newblock \emph{IEEE Access}, 7:\penalty0 80849--80860, 2019.

\bibitem[Hsu and Tu(2020)]{dual}
Ching-Sheng Hsu and Shu-Fen Tu.
\newblock Enhancing the robustness of image watermarking against cropping
  attacks with dual watermarks.
\newblock \emph{Multimedia Tools and Applications}, 79\penalty0 (17):\penalty0
  11297--11323, May 2020.
\newblock ISSN 1573-7721.
\newblock \doi{10.1007/s11042-019-08367-6}.
\newblock URL \url{https://doi.org/10.1007/s11042-019-08367-6}.

\bibitem[Zhu et~al.(2018)Zhu, Kaplan, Johnson, and Fei-Fei]{Zhu_2018_ECCV}
Jiren Zhu, Russell Kaplan, Justin Johnson, and Li~Fei-Fei.
\newblock {HiDDeN: Hiding Data with Deep Networks}.
\newblock In \emph{The European Conference on Computer Vision (ECCV)}, Sep
  2018.

\bibitem[{Wen} and {Aydore}(2019)]{wen2019romark}
Bingyang {Wen} and Sergul {Aydore}.
\newblock {ROMark: A Robust Watermarking System Using Adversarial Training}.
\newblock \emph{arXiv e-prints}, art. arXiv:1910.01221, October 2019.

\bibitem[{Huang} et~al.(2019){Huang}, {Niu}, {Guan}, and {Zhang}]{8673925}
Y.~{Huang}, B.~{Niu}, H.~{Guan}, and S.~{Zhang}.
\newblock Enhancing image watermarking with adaptive embedding parameter and
  psnr guarantee.
\newblock \emph{IEEE Transactions on Multimedia}, 21\penalty0 (10):\penalty0
  2447--2460, 2019.

\bibitem[{Luo} et~al.(2020){Luo}, {Zhan}, {Chang}, {Yang}, and
  {Milanfar}]{luo2020distortion}
Xiyang {Luo}, Ruohan {Zhan}, Huiwen {Chang}, Feng {Yang}, and Peyman
  {Milanfar}.
\newblock {Distortion Agnostic Deep Watermarking}.
\newblock \emph{arXiv e-prints}, art. arXiv:2001.04580, January 2020.

\bibitem[{Plata} and {Syga}(2020)]{spatial}
Marcin {Plata} and Piotr {Syga}.
\newblock {Robust Spatial-spread Deep Neural Image Watermarking}.
\newblock \emph{arXiv e-prints}, art. arXiv:, March 2020.

\bibitem[Ahmadi et~al.(2020)Ahmadi, Norouzi, Karimi, Samavi, and
  Emami]{ReDMark}
Mahdi Ahmadi, Alireza Norouzi, Nader Karimi, Shadrokh Samavi, and Ali Emami.
\newblock Redmark: Framework for residual diffusion watermarking based on deep
  networks.
\newblock \emph{Expert Systems with Applications}, 146:\penalty0 113157, 2020.
\newblock ISSN 0957-4174.
\newblock \doi{https://doi.org/10.1016/j.eswa.2019.113157}.
\newblock URL
  \url{http://www.sciencedirect.com/science/article/pii/S0957417419308759}.

\bibitem[Hamamoto and Kawamura(2020)]{JpegRotate}
Ippei Hamamoto and Masaki Kawamura.
\newblock Neural watermarking method including an attack simulator against
  rotation and compression attacks.
\newblock \emph{IEICE Transactions on Information and Systems},
  E103.D:\penalty0 33--41, January 2020.
\newblock \doi{10.1587/transinf.2019MUP0007}.

\bibitem[Goodfellow et~al.(2014)Goodfellow, Pouget-Abadie, Mirza, Xu,
  Warde-Farley, Ozair, Courville, and Bengio]{gan}
Ian Goodfellow, Jean Pouget-Abadie, Mehdi Mirza, Bing Xu, David Warde-Farley,
  Sherjil Ozair, Aaron Courville, and Yoshua Bengio.
\newblock Generative adversarial nets.
\newblock In \emph{Advances in Neural Information Processing Systems 27}, pages
  2672--2680, 2014.
\newblock URL
  \url{http://papers.nips.cc/paper/5423-generative-adversarial-nets.pdf}.

\bibitem[Hayes and Danezis(2017)]{steggan}
Jamie Hayes and George Danezis.
\newblock Generating steganographic images via adversarial training.
\newblock In \emph{Advances in Neural Information Processing Systems 30}, pages
  1954--1963, 2017.
\newblock URL
  \url{http://papers.nips.cc/paper/6791-generating-steganographic-images-via-adversarial-training.pdf}.

\bibitem[{Friend MTS}(2018)]{mts}
{Friend MTS}.
\newblock Comparing subscriber watermarking technologies for premium pay tv
  content.
\newblock Sep 2018.
\newblock URL
  \url{https://www.friendmts.com/wp-content/uploads/2018/09/FMTS_Watermarking_whitepaper_12pp_AW_update_WEB.pdf}.

\bibitem[Lin et~al.(2014)Lin, Maire, Belongie, Hays, Perona, Ramanan,
  Doll{\'a}r, and Zitnick]{coco}
Tsung-Yi Lin, Michael Maire, Serge Belongie, James Hays, Pietro Perona, Deva
  Ramanan, Piotr Doll{\'a}r, and C.~Lawrence Zitnick.
\newblock Microsoft coco: Common objects in context.
\newblock In David Fleet, Tomas Pajdla, Bernt Schiele, and Tinne Tuytelaars,
  editors, \emph{Computer Vision -- ECCV 2014}, pages 740--755, Cham, 2014.
  Springer International Publishing.
\newblock ISBN 978-3-319-10602-1.

\bibitem[Kingma and Ba(2014)]{adam}
Diederik Kingma and Jimmy Ba.
\newblock {Adam: A Method for Stochastic Optimization}.
\newblock \emph{International Conference on Learning Representations}, Dec
  2014.

\end{thebibliography}

\end{document}